\documentclass[a4,12]{article}

\usepackage{graphicx}
\usepackage{setspace} 
\usepackage{amssymb}
\usepackage{color}
\usepackage{amsthm}
\usepackage{lineno}
\usepackage{ifpdf}
\usepackage[english]{babel}
\usepackage{color}
\usepackage[T1]{fontenc}
\usepackage[latin1]{inputenc}
\usepackage{graphicx}
\usepackage{latexsym}
\usepackage{epsfig}
\usepackage{graphicx}
\usepackage{amsmath}
\usepackage{amsfonts}
\usepackage{amssymb}
\usepackage{dcolumn}
\usepackage{color}
\usepackage{xcolor}
\usepackage{ulsy}
\input{epsf}
\usepackage{ifpdf}
\usepackage[stable]{footmisc}
\usepackage[comma,authoryear]{natbib} 

\makeatother

\ifpdf
  \usepackage[pdftex]{hyperref}
\else
  \usepackage[hypertex]{hyperref}
\fi

\hypersetup{
  a4paper,
  colorlinks=true,
  urlcolor=red,
  raiselinks=true,
  linkcolor=red, 
  runcolor =green,
  citecolor=blue,
  pdfpagemode=none,
  pdftitle={Paper Overshoot},
  pdfauthor={Teodor Burghelea},
  pdfcreator={$ $ Teodor I. Burghelea$ $},
  pdfsubject={viscosity overshoot},
  bookmarks=true
  pdfkeywords={}
}
\begin{document}

\title{On the "viscosity maximum" during the uniaxial extension of a low density polyethylene}

\author{Teodor I. Burghelea, Zden\v{e}k Star\'{y}, Helmut M\"{u}nstedt\\
Institute of Polymer Materials, Friedrich- Alexander University,\\ Martensstrasse 7, D-91058 Erlangen}

%\affiliation{Institute of Polymer Materials, Friedrich-Alexander University, Martensstrasse 7, D-91058 Erlangen}
\maketitle

\begin{abstract}
An experimental investigation of the viscosity overshoot phenomenon observed during uniaxial extension of a low density polyethylene is presented. For this purpose, traditional integral viscosity measurements on a M\"{u’}nstedt type extensional rheometer are combined with local measurements based on the in-situ visualization of the sample under extension.  For elongational experiments at constant strain rates within a wide range of Weissenberg numbers (Wi), three distinct deformation regimes are identified. Corresponding to low values of Wi (regime I), the tensile stress displays a broad maximum. This maximum can be explained by simple mathematical arguments as a result of low deformation rates and it should not be confused with the viscosity overshoot phenomenon.   
Corresponding to intermediate values of Wi (regime II), a local maximum of the integral extensional viscosity is systematically observed. However, within this regime, the local viscosity measurements reveal no maximum, but a plateau. Careful inspection of the images of samples within this regime shows that, corresponding to the maximum of the integral viscosity, secondary necks develop along the sample. The emergence of a maximum of the integral elongational viscosity is thus related to the distinct inhomogeneity of deformation states and is not related to the rheological properties of the material. In  the fast stretching limit (high Wi, regime III), the overall geometric uniformity of the sample is well preserved, no secondary necks are observed and both the integral and the local transient elongational viscosity show no maximum. A detailed comparison of the experimental findings with results from literature is presented.  
    
\end{abstract}

%\begin{keyword}
%% keywords here, in the form: 
 %uniaxial extension \sep viscosity overshoot \sep M\"{u’}nstedt rheometer \sep homogeneity of deformation



%\end{keyword}

%\end{frontmatter}

%\linenumbers
%\tableofcontents
\section{Introduction} \label{sec_intro}
Uniaxial extension is the dominant flow in many industrial processes and, therefore, accurate measurements of the extensional rheological properties of polymer melts are very important. From a more fundamental point of view, reliable measurements of the rheological properties in extension are crucial for validating existing theoretical models and suggesting new approaches. In spite of the universally recognized need for reliable elongational measurements of polymer melts, the development of extensional rheometric equipment has progressed slowly during the past three decades. A reliable design of an extensional rheometer has met several practical difficulties and one of the taughest is to generate a homogeneous extensional flow. 
Several techniques to measure the elongational properties of polymer melts have been proposed: the Rheometrics Melt Extensiometer (RME) by Meissner \cite{meissner1}, the supporting oil bath design by M\"{u}nstedt \cite{m4}, and the SER by Sentmanat \cite{sentamat1,  sentamat2}.
 A comprehensive review of these different approaches to extensional rheology of polymer melts is beyond the scope of this investigation and can be found in \cite{schweizer} and more recently in \cite{handbook}. We note that for each of these approaches the homogeneity of deformation states is crucial for reliably assessing the elongational properties of the material. Though previously recognized by most experimentalists, it is our belief that this issue did not receive the proper attention. Only very recently the true danger of sample non-uniformity during elongation has been made explicit, by measuring locally both the stresses and the strain and showing that in the case of strongly nonuniform samples the classical extensional measurements become completely unreliable, \cite{localelongational1}. In \cite{localelongational1} it has been demonstrated by combined traditional integral viscosity measurements and local viscosity measurements based on in-situ local measurements of the sample diameter that geometric non-uniformities of the sample under elongation typically result in completely unreliable viscosity data. Moreover, it has been shown that even initially homogeneous (perfectly cylindrical) samples loose their uniformity at high enough Hencky strains and thus, the impact of sample non-homogeneity on the viscosity measurements is always an issue to worry about during extensional tests. Also very recently, a full numerical simulation of the extension process in a $SER$ rheometer demonstrated that the loss of sample homogeneity during deformation leads to a strong strain localization along the sample which ultimately translates into unreliable measurements of the transient elongational viscosity \cite{hassager2009, hassager2009PRL}.

Here a more elaborated version of the method proposed in \cite{localelongational1} is employed to investigate a long standing problem in the extensional rheology of polymer melts, the viscosity (or tensile stress) overshoot observed during the uniaxial extension of some polymer melts at a constant rate of deformation. 

Since the early days of extensional rheometry it has been observed that some strain hardening materials under uniaxial extension display a clear maximum in the transient extensional viscosity right before (typically within less than a Hencky strain unit) the physical rupture of the sample \cite{m2, meissnerovershoot, meissner1, nielsen1}. There is only one experimental paper we are aware of \cite{rasmussen}, which also presents an extended (over several Hencky strain units prior to the physical rupture of the sample) plateau after the stress maximum. However, the authors of this study present no experimental evidence on the homogeneity of the sample during the elongation process. This local maximum in the tensile stress, followed or not by a plateau, has been coined as \textit{"viscosity overshoot"}. 
The existence of a true viscosity overshoot is important from both a practical and a fundamental point of view. 
In many processing and industrial settings it is important to know whether a true steady state behaviour can be reached under extension at a constant rate and if not, to understand how this fact influences the physical rupture of the material. 
From a theoretical point of view, in our opinion, this phenomenon is not yet fully understood. The POM-POM model for branched polymer melts \cite{mcleish} and the molecular stress function (MSF) model \cite{wagner4}  predict a monotone increase of the transient extensional viscosity. Other theoretical works, however, are able to predict an overshoot in viscosity, \cite{wagnerovershoot, wagner3}. Even more worrying, recent theoretical models seem to be able to fit both a maximum and/or a steady state of the transient extensional viscosity \cite{wagner3}. This simply means that the phenomenology behind the stress maximum/overshoot remains elusive.

The implications of the stress overshoot phenomenon during uniaxial extension are, in our opinion, even more important. Recently, based on the observation of the stress overshoot phenomenon in both uniaxial extension of polymer melts \cite{wang1, wang3} and startup shear of entangled polymer solutions \cite{wang4}, a universality claim \textit{"Entangled liquids are solids"} has been very recently formulated \cite{wang2}.   
Though we do understand how important and appealing a universal behaviour is and we do accept that some similarities between polymer melts under elongation and entangled solutions in startup shear may exist, we believe that the claim above should be still considered very cautiously, at least because of the reasons below: 

\begin{enumerate}

\item The term "overshoot" implies, to our best understanding, a local maximum followed by a plateau corresponding to lower stress values. Whereas such plateau has been observed for entangled solutions \cite{wang4}, the data concerning polymer melts presented in \cite{wang1, wang3} display only a maximum but no plateau: the sample breaks just after the maximum. A true overshoot behavior (a maximum followed by a plateau) has been observed in \cite{rasmussen} but at much higher Hencky strains. 

\item The stress overshoot during the uniaxial extension of entangled polymer melts reported in \cite{wang1, wang3} refers to the so-called "engineering stress" which is not a real stress but just a tensile force normalized by a constant (the initial area of the sample under investigation). The (physical) true stress (which is calculated by dividing the tensile force by the actual cross section of the sample) does not always exhibit an overshoot behaviour and if it does (e.g. for the strain hardening materials at high enough rates of deformation) this occurs at significantly larger Hencky strains. 
\end{enumerate} 
In the view of the remarks above, we believe that before an analogy between the stress maximum (and only rarely a true stress overshoot, \cite{rasmussen}) observed for polymer melts under extension and the stress overshoot observed for entangled polymer solutions in startup shear is stated, a deeper understanding of each phenomenon is needed. 

As the elongational viscosity overshoot phenomenon has been observed at high Hencky strains, prior to the physical rupture of the sample,  a proper understanding of this phenomenon might also shed light on different mechanisms of failure during elongation, which remains an elusive goal \cite{denn2003, denn2004}.

\section{Description of the experiments} \label{sec_experimental}

\subsection{Experimental apparatus and techniques} \label{subsec_apparatus}

\begin{figure*}
\begin{center}
\centering
\includegraphics[width=12cm]{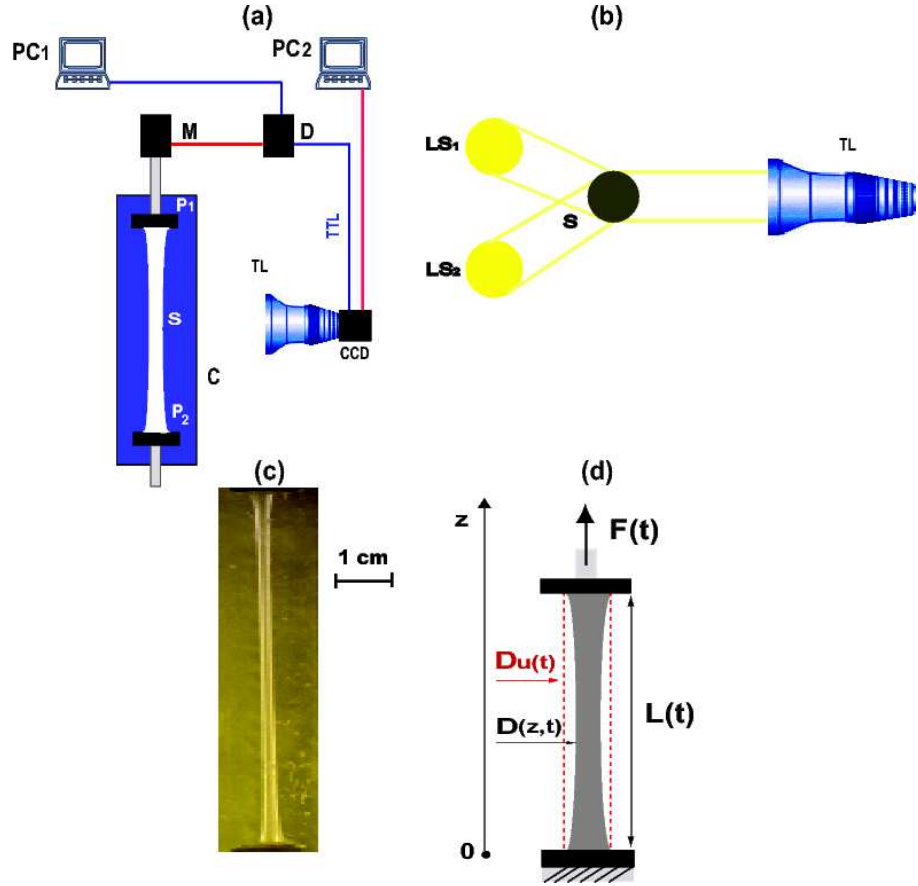}
\caption{(a) Schematic view of the experimental apparatus: \textbf{C}- oil bath, $\mathbf{P_{1}}$ and $\mathbf{P_{2}}$ - top and bottom plates of the rheometer, \textbf{S}- the sample under investigation, \textbf{M}- AC servo motor, \textbf{D}- the control drive of the rheometer, $\mathbf{PC_{1,2}}$ - personal computers, \textbf{TL}- telecentric lens, \textbf{CCD}- video camera. (b) Sample illumination and imaging: $\mathbf{LS_{1}}$ and $\mathbf{LS_{2}}$- linear light sources, \textbf{S}- the sample under investigation. (c)  Example of a telecentric sample image corresponding to $\epsilon_{H}=2.7$. The field of view was actually larger but the image has been cropped for clarity reasons.  (d) Principle of the local measurements of the extensional viscosity. The vertical dotted lines represent the contour of an ideal uniform sample.} \label{f1}
\end{center}
\end{figure*}

The experiments have been conducted with a M\"{u}nstedt type extensional rheometer built in the house which is illustrated in Fig. \ref{f1}(a). 
A detailed description of this device can be found elsewhere, \cite{m1}. The specimen \textbf{S} under investigation is clamped between the plates $\mathbf{P_{1}}$ and $\mathbf{P_{2}}$ of the rheometer and immersed in a silicone oil bath \textbf{C} to minimize gravity and buoyancy effects, 
Fig. \ref{f1}(a).

While the bottom plate $\bf{P_2}$ is stationary, the top plate $\bf{P_1}$ is moved vertically by an AC-servo motor \textbf{M}, 
controlled by an analogue to digital converter installed on the computer $\bf{PC_1}$ . The sample is illuminated from behind by to linear light sources
$\bf{LS_{1}}$ and $\bf{LS_{2}}$ disposed as shown in the schematic top view presented in Fig. \ref{f1} (b). The idea behind the back-light illumination arrangement is to obtain a maximum of brightness only on the edges of the sample and thus to allow accurate identification of the sample edges and reliably measure its diameter. A major difficulty in imaging a considerably elongated sample comes from the high aspect ratio (height to width) of the corresponding field of view, which during extensional experiments at large Hencky strains may be as large as $1:50$. If a regular entocentric lens (with the entrance pupil located inside the lens) is used both the resolution and the level of geometrical distortion are unsatisfactory for high accuracy measurements of the sample diameter. Additionally, corresponding to large Hencky strains, both the frame brightness and the degree of focusing become uneven through the field of view if the sample is imaged in divergent light. To circumvent these problems, we use in our study a high resolution telecentric lens with the entrance pupil located at infinity, (VisionMes 225/11/0.1, Carl Zeiss) which images the sample in parallel light and delivers frames with very uniform brightness and free of distortions (geometrical aberrations), perspective errors and edge position uncertainties. A typical image of the sample is presented in Fig. \ref{f1} (c).
Images of the sample under elongation are acquired in real time using a high resolution (3000 by 1400 pixels full frame, which translates into roughly $60 ~\mu m$ spatial resolution) low noise camera (Pixelink from Edmunds Optics) at a speed of $3$ frames per second. The video camera is installed on a second computer, $\bf{PC_2}$. 
The image acquisition is digitally synchronized with the rheometer via a transistor-transistor logic ($TTL$) trigger signal sent by the rheometer drive \textbf{D} to the camera. 

\subsection{Materials and their rheological properties} \label{subsec_materials}
The material used in this study is a low-density polyethylene from Lyondell Bassel with the trade name Lupolen 1840 D. Several molecular and rheological characteristics of the material are summarized in Table \ref{tabel_rheo}.
LDPE 1840 D has a branched molecular structure which has been systematically characterized by Nordmaier and co-workers \cite{nordmaier1, nordmaier2}. It has a broad molar mass distribution with a rather large molar mass, $M_w$, and a pronounced high molar mass tail. 

\begin{table}
\begin{center}
\begin{tabular}{|c|c|c|c|c|}
\hline  $\mathbf{M_w~(kg/mol)}$ & $\mathbf{M_w/M_n}$ & $\mathbf{\eta_0~(Pas)}$ & $\mathbf{J_e^0~(10^{-4}Pa^{-1})}$ & $\mathbf{\lambda~(s)}$ \\ 
\hline
\hline 377 & 18 & 833000 & 13.5 &1100 \\ 
\hline 
\end{tabular} 
\end{center}
\caption{Molecular and rheological characteristics of Lupolen 1840 D at $T=140 ^\circ C$.}\label{tabel_rheo}
\end{table}

The influence of the broadly distributed molar mass and the branched molecular structure of the material on its rheological properties in shear has been recently investigated experimentally \cite{resch1}. Due to the broadly distributed molar mass and the degree of chain branching, the maximum relaxation time of the material, $\lambda$, is quite high. It is calculated as $\lambda=J_e^0 \cdot \eta_0$, where $J_e^0$ and $\eta_0$ are the steady 
-state recoverable compliance and the zero shear viscosity, respectively.

\subsection{Preparation of the samples}\label{subsec_preparation}
For elongational measurements in the M\"{u}nstedt rheometer cylindrical specimens were used. At first a strand was extruded through a capillary at $190 ~^\circ C$ using a piston extrusion machine. The diameter $D$ of the die was $4.6~mm$, the length $L=18.4~mm$, and the apparent shear stress applied was $43.7~ kPa$. These extrusion conditions give rise to a strand with a diameter of about $8 ~mm$ after annealing. The strand was extruded into a vessel containing an ethanol-water mixture ($90/10 ~vol. ~\%$) in order to ensure homogeneous strand diameter along the axes of the extrusion. This procedure leads to specimens with a relative deviation of a diameter smaller than $2 \%$.
After extrusion the strand was annealed in a silicone oil bath at $150 ~^\circ C$ for $20~ min$. This step ensures a complete relaxation of the thermal stress accumulated within the sample, which is necessary in order to suppress any stress history effects and obtain reliable and accurate rheological data. 
The initial length and diameter of each sample were $D_0=8~mm$ and $L_0=5~mm$, respectively.
The specimen's surface was etched by air plasma at room temperature in order to increase its surface energy. The samples were glued to aluminium clamps using a two-component epoxy resin adhesive, Technicoll $8266/67$. These clamps serve to fix the specimen to the pulling rod and the force transducer of the rheometer. At last, the specimens were kept in an oven at $80 ^\circ~ C$ for $2$ hours for a complete curing of the glue.

\subsection{Data analysis} \label{subsec_anlysis}
The first step of our data analysis procedure was to interpolate both the image sequence and the data acquired by the M\"{u}nstedt rheometer on a common time axis, so a direct comparison between the integral viscosity measurements and the shape of the sample under deformation can be made.
Prior to analysis, each image has been compensated for non uniform brightness using a standard adaptive histogram equalization algorithm implemented under Matlab. By identifying the edges of the sample from each image, the distribution of diameters along the actual length of the sample is measured. This allows the calculation of the stress distribution along the sample. The true tensile stress corresponding to each deformation state is defined by the mean of the stress distribution and the error bars are defined by the root mean square deviation (rms) of the stress distribution.

\section{Results}
\subsection{On the relation between a maximum in the tensile stress and sample uniformity}\label{subsec_analytical_condition}

Previous theoretical (\cite{wagnerovershoot, wagner3}) and experimental (\cite{rasmussen}) studies  aimed high by attempting to explain a stress maximum in terms of the molecular scale dynamics during extension but very little questioned the reliability of the existing extensional data (particularly in relation with the sample homogeneity during extension). We would like to address in the following a more modest question which is decisive for a fundamental understanding of the elongational behaviour of polymer melts and its theoretical description: \textit{Is a maximum of the transient tensile stress compatible with a uniform deformation process?}. By uniform deformation we understand a uniaxial deformation at a constant rate, $\dot{\epsilon}$, for which the diameter of the sample, $D(t)$, is constant along the actual length of the sample $L(t)$ (see Fig. \ref{f1}(d)) and, based on the incompressibility condition, is given by $D_u(t)=D_0~e^{-\dot{\epsilon}t/2}$. The transient stress $\sigma(t)$ during the extension of a real sample with a coordinate-dependent diameter $D(z,t)$, Fig. \ref{f1}(d),  can be written as an average of the local stresses along the direction of extension, $z$: 

\begin{equation}
\sigma(t)=\frac{4 F(t)}{\pi L(t)} \int_0^{L(t)} {\frac{1}{D^2(z,t)}dz}\label{eq_stressaverage}
\end{equation}
Here $F(t)$ stands for the tensile force which is coordinate independent. By a simple algebraic manipulation using Leibniz's theorem one can easily show that, corresponding to a stress maximum ($\frac{d\sigma (t)}{dt}=0$), the following condition should be fulfilled:
\begin{equation}
\left\lbrace  \frac{\pi \sigma(t)}{4} - \frac{F(t)}{D^2[L(t),t]} \right\rbrace \frac{dL(t)}{dt} =\frac{1}{L(t)}\int_0^{L(t)} \frac{\partial \left( \frac{F(t)}{D^2[z,t]} \right)}{\partial t}dz\label{eq_condition}
\end{equation}
One can notice that the curly bracket on the left hand side of the equation above is nothing but a measure of the sample uniformity. Indeed, if one imposes the condition that the sample preserves its cylindrical shape at all times ($D[z,t]=D[0,t]=D[L(t),t]=D_u(t), \forall z\in [0, L(t)]$), the left hand side of the equation above vanishes and the stress maximum condition reduces to $\frac{d \left( \frac{F(t)}{D_u^2(t)} \right)}{d t}=0$. Thus, if the sample is assumed to deform uniformly (the curly bracket in the left hand side of Eq. \ref{eq_condition} vanishes), corresponding to a stress maximum the tensile force should scale exponentially, $F(t) \propto e^{-\dot{\epsilon}t}$. However, we point out that the tensile force may scale nearly exponentially if the rates of deformation are very small (note that the $dL(t)/dt \propto \dot{\epsilon}$) even in the case that the deformation process is non uniform (the curly bracket in Eq. \ref{eq_condition} is non zero). 
To conclude, if the deformation process is assumed to be homogeneous and if a local maximum in stress is attained then, in the neighbourhood of the stress maximum, the tensile force should scale as an exponentially decaying function with a rate equal to the rate of elongation, $\dot {\epsilon}$.
The validity of the analytical condition for a maximum in stress  is discussed in the next subsection, \ref{subsec_extensionregimes}. 

\subsection{The transient tensile force and tensile stress in different regimes of extension}\label{subsec_extensionregimes} 

\begin{figure*}[h]
\begin{center}
\centering
\includegraphics[width=10cm]{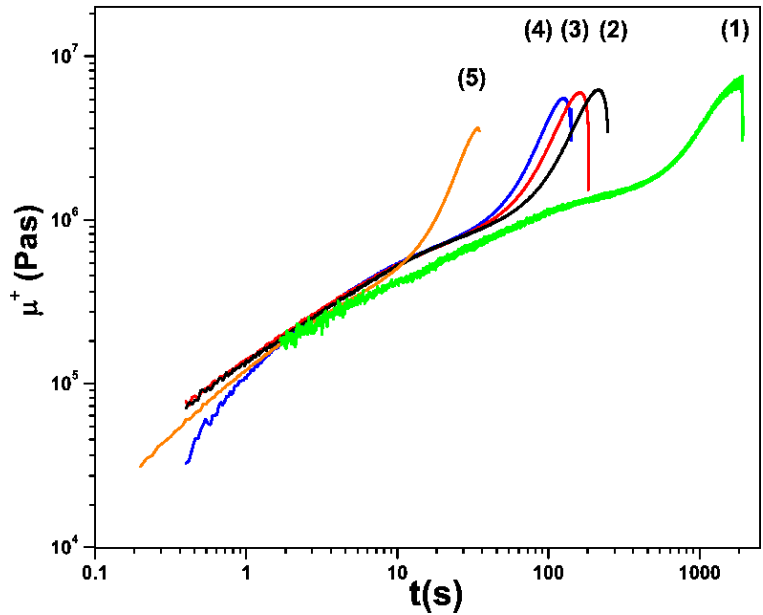}
\caption{Transient elongational viscosities at various rates of deformation: (1)- $\dot{\epsilon= 0.002 s^{-1}}$, (2)- $\dot{\epsilon=0.015 s^{-1}}$, (3)- $\dot{\epsilon=0.02 s^{-1}}$, (4)- $\dot{\epsilon= 0.025 s^{-1}}$, (5)- $\dot{\epsilon=0.09 s^{-1}}$. Each data set has been acquired until the physical rupture of the sample occurred.} \label{moreetas}
\end{center}
\end{figure*}

Integral measurements of the transient elongational viscosity are presented in Fig. \ref{moreetas}. The integral viscosity is obtained by measuring the transient tensile force, $F(t)$, using the assumption that the diameter of the sample is independent on the vertical coordinate, $D(z,t)=D_u(t)$.  
Each data set has been acquired until the physical rupture of the sample occurred. Except for the linear range of deformation, $\epsilon_H <1$, the shape of the transient elongational viscosity depends considerably on the (constant) rate at which the material is deformed, $\dot{\epsilon}$. Thus, depending on the rate of deformation, the integral transient viscosity may display either a clear maximum (curves 1-4, Fig. \ref{moreetas})or a monotonic increase (curve (5), Fig. \ref{moreetas}). As clearly suggested by Eq. \ref{eq_condition} and the discussion presented in Sec. \ref{subsec_analytical_condition}, in order to understand the physical reasons underlying the viscosity maximum visible for the curves (1-4), one has to focus not only on the tensile stress but on the tensile force as well.     

\begin{figure*}
\begin{center}
\centering
\includegraphics[width=14cm]{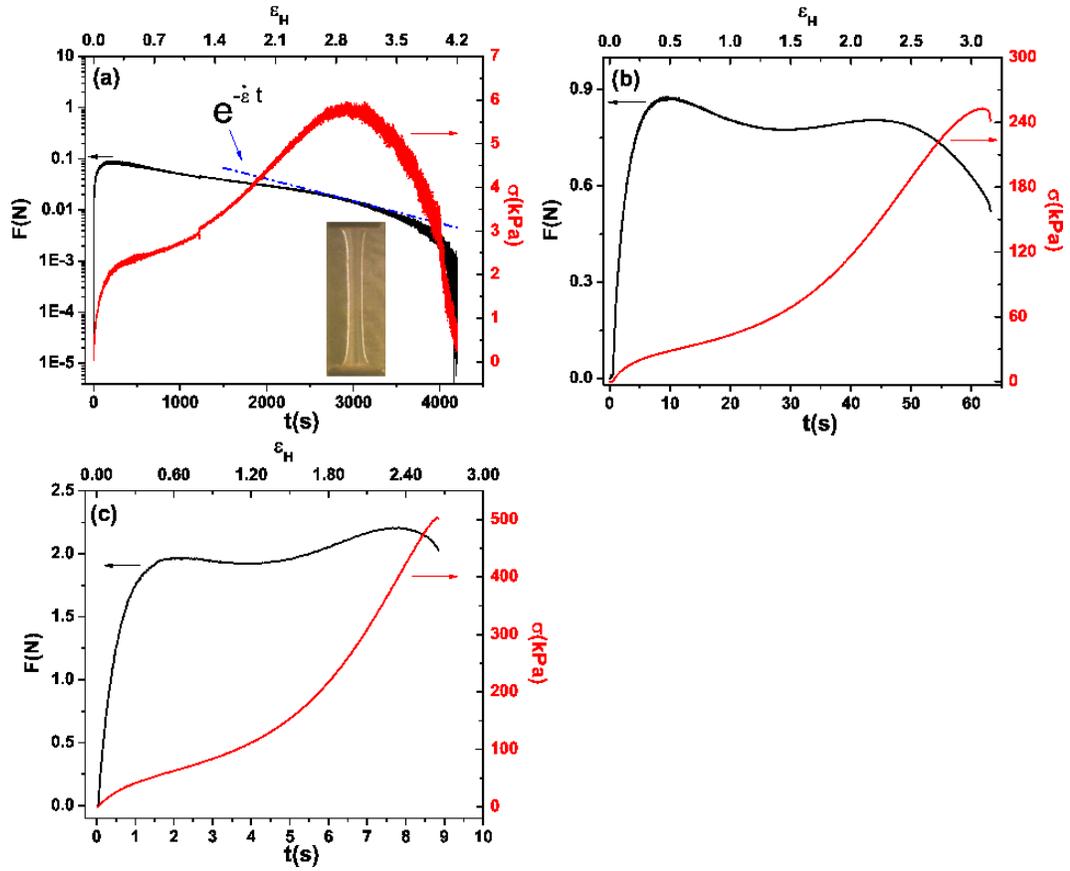}
\caption{Transient tensile forces and stresses corresponding to different regimes of extension: (a) $\dot{\epsilon} =0.001 s^{-1} (Wi=1.1)$ ,
(b) $\dot{\epsilon} =0.05 s^{-1} (Wi=55)$, (b) $\dot{\epsilon} =0.3 s^{-1} (Wi=330)$.
The inset in panel (a) displays the image of the sample corresponding to the stress maximum.
} \label{forces}
\end{center}
\end{figure*}

In Fig. \ref{forces} the transient tensile forces and stresses measured for three different values of the Weissenberg number, $Wi$, are presented. The Weissenberg number is defined as $Wi=\dot \epsilon \cdot \lambda$. 
Corresponding to $Wi=1.1$, the tensile stress displays a broad maximum, Fig. \ref{forces} (a).

In order to connect the emergence of the stress maximum with the discussion presented in \ref{subsec_analytical_condition}, we need to discuss the homogeneity of the sample around the stress maximum. The force maximum visible in Fig. \ref{forces} (a) (which corresponds to a shoulder in the transient tensile stresses) represents the onset of a primary non-uniformity of the specimen, as predicted by the Consid\`{e}re criterion, \cite{considere}. Such geometric non-uniformity of the sample (initially localized near the plates $\mathbf{P_{1,2}}$ of the rheometer) occurs in most of the extensional experiments and it is related to the rigid boundary conditions near the clamping points of the sample under investigation. Thus, the emergence of this effect depends little on the molecular structure of the material: it can be observed for rubbers, for linear polymer melts and even during cold drawing experiments (Ref. \cite{strobl} and the references therein). With increasing time the tensile stress reaches a maximum around $\epsilon_H \approx 3$. 
 
It is interesting to note that in the neighbourhood of the stress maximum the tensile force scales exponentially, in agreement with the derivation presented in Sec. \ref{subsec_analytical_condition}. This fact deserves a brief discussion.
As shown in Sec. \ref{subsec_analytical_condition}, a nearly exponential scaling of the tensile force around the stress maximum can be found either when the sample deforms uniformly or when the rates of deformation are very small. Both these situations make the left hand side of Eq. \ref{eq_condition} very small allowing one to solve it for a nearly exponential tensile force.  
Corresponding to the Hencky strain where the local stress maximum is observed, the deformation of the sample is not homogeneous, as illustrated in the inset in Fig. \ref{forces} (a). Therefore, the nearly exponential scaling of the tensile force is due to the smallness of the deformation rate $\dot \epsilon =0.001~s^{-1}$ solely. This broad stress maximum observed at very low $Wi$ should be not be confused with the viscosity overshoot phenomenon which was observed in a faster regime of stretching where significant strain hardening effects were present, \cite{wagnerovershoot, rasmussen}. 

A stress maximum is clearly observed at the larger rate of deformation $\dot{\epsilon} =0.05 s^{-1} (Wi=55)$, Fig. \ref{forces} (b). We note that we do not observe a true stress overshoot in the sense that the local stress maximum is not followed by a plateau. Based on the derivation presented in Sec. \ref{subsec_analytical_condition}, one can easily conclude that, corresponding to the local maximum of the tensile stress, the deformation is inhomogeneous. Indeed, as shown in Sec. \ref{subsec_analytical_condition} if a homogeneous deformation is assumed then, corresponding to the stress maximum, the tensile force should decay exponentially. This is clearly not the case for the data presented in Fig. \ref{forces} (b).
Corresponding to $\dot{\epsilon} =0.3 s^{-1} (Wi=330)$, a local maximum of the tensile stress is no longer observed: the sample breaks before the tensile stress reaches either a maximum or a steady state.

In order to get a more complete picture of how the shape of the transient tensile force/stress is influenced by the forcing conditions and identify the deformation regime where a stress maximum is observed, measurements similar to those presented in Fig. \ref{forces}(a-c) were performed in a wide range of Weissenberg numbers, spanning nearly three decades.
\begin{figure*}
\begin{center}
\centering
\includegraphics[width=10cm]{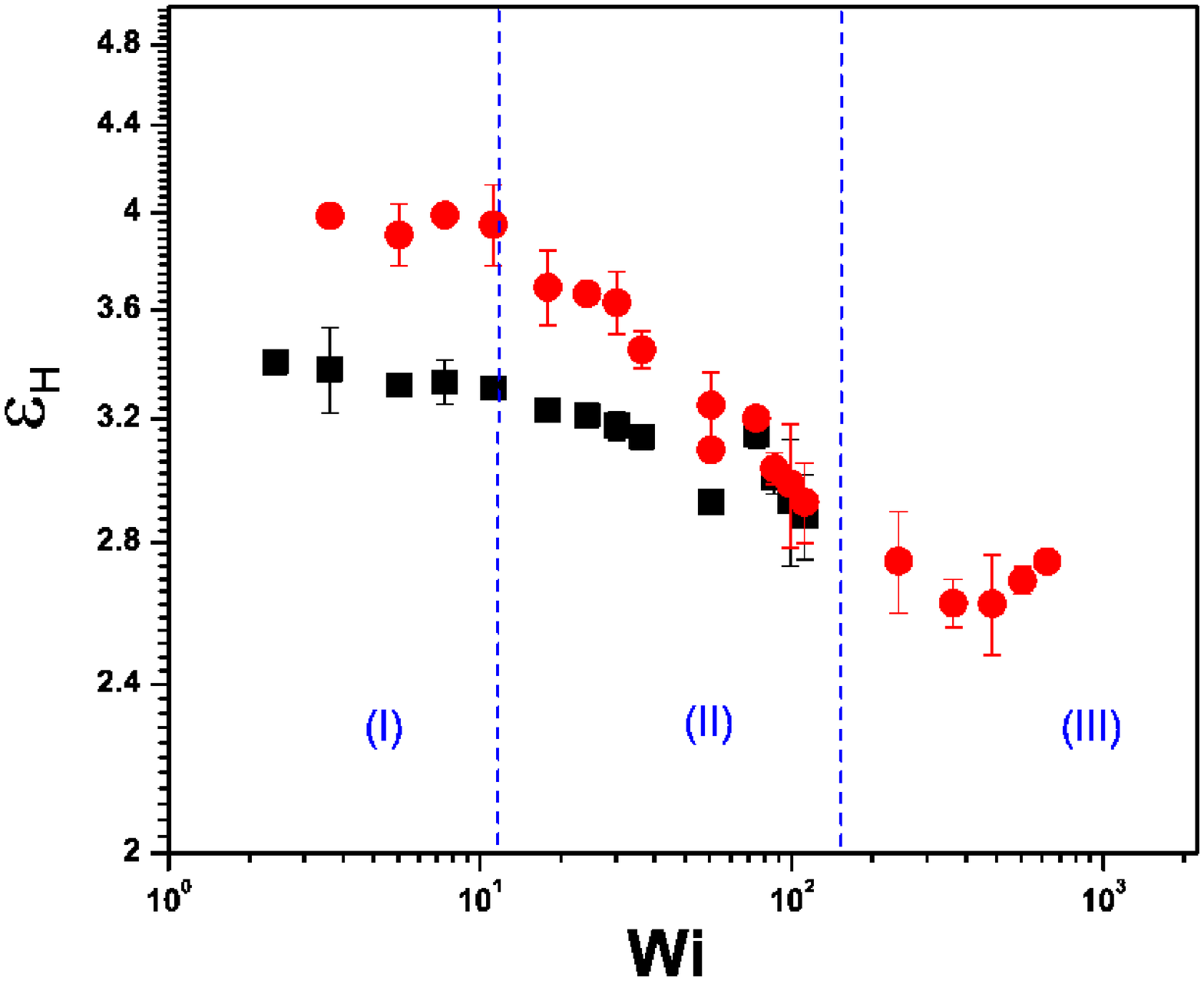}
\caption{Dependence of the Hencky strain corresponding to the stress maximum (squares) and physical rupture of the sample (circles) on the Weisenberg number, $Wi$. The vertical dotted lines delineate the extension regimes (I), (II) and (III).} \label{phasediagram}
\end{center}
\end{figure*}

The results of these measurements are summarized in Fig. \ref{phasediagram} which presents the $Wi$ dependence of the Hencky strains corresponding to a maximum in the tensile stress (the squares) and to the physical rupture of the sample (the circles). The Hencky strains presented in Fig. \ref{phasediagram} have been identified using integral measurements of the transient tensile force and stress, similar to those presented in Fig. \ref{forces}.  
 
Based on a careful inspection of the dependencies $F=F(t)$ and $\sigma=\sigma(t)$, three different regimes of extension can be distinguished, Fig. \ref{phasediagram}. For $Wi \leq 10$ the Hencky strains corresponding to a stress maximum and to the physical rupture of the sample are practically independent on $Wi$. In the deformation regime (I), the maximum of the tensile stress and the physical rupture of the sample occur at a nearly constant Hencky strain each and they are separated by (roughly) $0.8$ Hencky strain units. We once more emphasize that the broad stress maximum observed within this regime can be explained as a result of an inhomogeneous deformation process at small deformation rates and is not necessarily related to the molecular structure of the material \footnote{Such a broad stress maximum has been observed at the \textit{Institute of Polymer Materials} for polystyrene melts and several polymer blends as well during elongational experiments at low rates of deformation.}.
 
As the rate of deformation is increased ($Wi>10$) a second deformation regime is observed. The Hencky strains corresponding to a stress maximum and to the physical rupture of the sample depend significantly on the Weissenberg number and they get progressively closer to each other as $Wi$ is increased. This finding suggests that the emergence of a stress maximum and the physical rupture of the sample are interconnected phenomena. An argument supporting this hypothesis is that, corresponding to the lower bound of the second deformation regime, (II), the time scale of the flow (estimated here as $\tau_f=1/ \dot{\epsilon} \approx 100~$) is significantly smaller than the largest relaxation time of the material, $\lambda$, given in Table \ref{tabel_rheo} as $1100~s$.   
Thus, the deformation state corresponding to a maximum in the tensile stress is likely to be "remembered" until the physical rupture of the sample occurs.  
Within the deformation regime (II), the transient elongational viscosity displays a local maximum which narrows as the rate of deformation increases.
Ultimately, if the Weissenberg number is increased even further, $Wi>150$, a third deformation regime, (III), is observed. Within this deformation a local maximum  of the tensile stress is no longer observed and the Hencky strain corresponding to the physical rupture of the sample becomes practically independent of $Wi$. 
 A more detailed characterization of the deformation regimes (I-III) in connection with different failure mechanisms will be presented elsewhere. 
This paper is mostly dedicated to the deformation regime (II) with a particular focus on the maximum of the transient tensile stress or extensional viscosity, respectively.

\subsection{Integral versus local measurements of the tensile stress}

Integral stress measurements rely heavily on the homogeneity of the deformation because they use the assumption that the diameter of the sample is coordinate-independent, $D(z,t)=D_u(t)$. As opposed to this, the local stress measurements using Eq. \ref{eq_stressaverage} can properly account for the deviations from a homogeneous deformation process.
In the following a direct comparison between integral and local measurements of the tensile stress is presented.

\begin{figure*}
\begin{center}
\centering
\includegraphics[width=7cm]{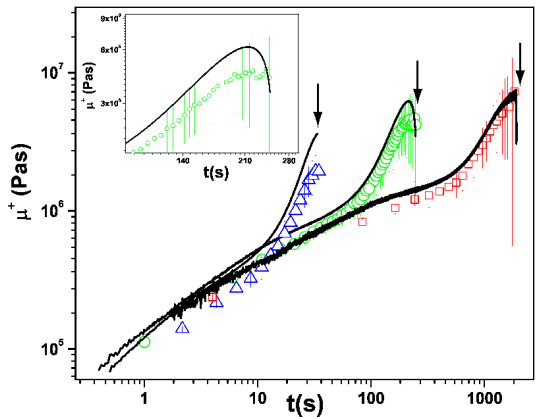}
\caption{Comparison between the integral (full lines) and the locally measured elongational viscosity in each of the deformation regimes presented in Fig. \ref{phasediagram}: squares - $Wi= 2.2$ (regime I), circles - $Wi=16.5$ (regime II), triangles - $Wi=300$ (regime III). The error bars are defined by the root mean square deviation of the local stresses along the sample. The inset presents a magnified view of the data acquired in regime II. The vertical arrows indicate the physical rupture of the sample.} \label{eta}
\end{center}
\end{figure*}

Results of such a comparison corresponding to each of the deformation regimes previously discussed are presented in Fig. \ref{eta}. One can note that, regardless the value of $Wi$, the integral transient viscosity lies systematically above the locally measured one even in a linear range of deformation, $\epsilon_{H} < 1$. It has recently been shown that this systematic overestimation of the transient extensional viscosity in the linear range is related to the large retardation times of LDPE 1840 D, \cite{localelongational2}. In the nonlinear range, the differences between integral and local measurements of the transient elongational viscosity seem to increase with the rate of deformation. Corresponding to the second deformation regime (II) ($Wi=16.5$) where the integral transient elongational viscosity displays a maximum, the locally measured elongational viscosity has no maximum but seems to reach a plateau, as clearly visible in the inset of Fig. \ref{eta}.  
This result has systematically been reproduced over the entire second regime of deformation (data not shown here), suggesting that maximum of the transient elongational viscosity might not be a true rheological feature of the material but merely an artefact related to the experimental procedure.
In the range of high $Wi$, (regime III) neither a maximum nor a plateau of the transient elongational viscosity is observed. Within this regime, the local and the true viscosity measurements agree qualitatively, though they are quantitatively different.

\subsection{Geometric non-uniformity of the sample and its relation with the stress maximum}

\begin{figure*}
\begin{center}
\centering
\includegraphics[width=15cm,height=11cm]{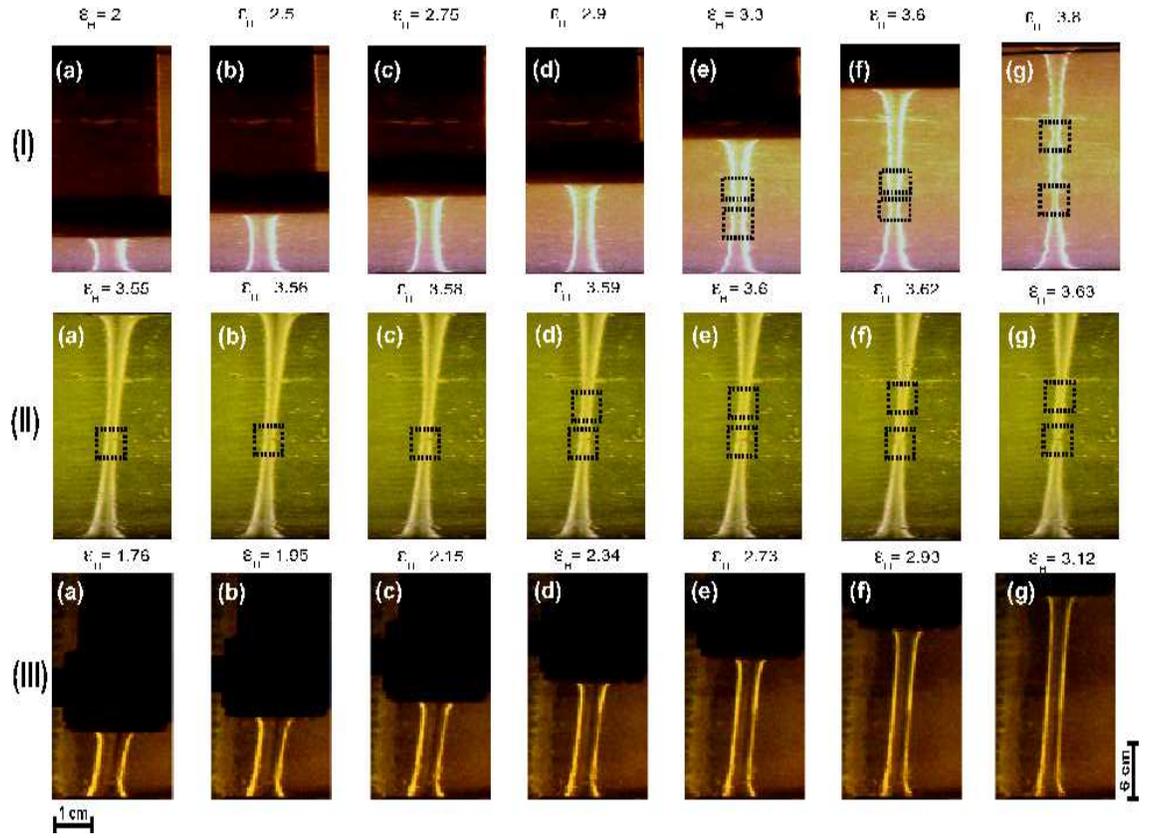}
\caption{Sequence of specimen images under deformation at different Wi. The image rows (from top to bottom) correspond to: $Wi= 2.2$ (region I), $Wi=16.5$ (region II), $Wi=99$ (region III). The aspect ratio of each image has been modified in order to enhance the clarity. The dotted squares indicate the location of the necks. The Hencky strains are indicated on the top of each image.} \label{frames}
\end{center}
\end{figure*}

In Fig. \ref{frames} we display images of the sample corresponding to each of the deformation regimes presented in Fig. \ref{phasediagram} and at several Hencky strains. The images corresponding to the highest Hencky strains (the last column in Fig. \ref{frames}) are the last images acquired prior to the physical rupture of the sample.  
The images presented in Fig. \ref{frames} have been rescaled in order to enhance the clarity of the presentation. However, this does not alter the main message concerning the geometric uniformity of the sample. 

Within the first regime of deformation (I) (first row from the top in Fig. \ref{frames}), the shape of the sample deviates strongly from a cylindrical one. The onset of these geometric non-uniformities (the primary neck extended over the entire length of the sample) occurs at low Hencky strains ($\epsilon_{H} \leq 1$) and, according to the Consid\`{e}re criterion, \cite{considere}, is related to the local maximum in the tensile force observed in Fig. \ref{forces}(a). In a range of high Hencky strains ($\epsilon_H \approx 3.5$) prior to the physical rupture of the sample, secondary necks develop in the proximity of the midpoint of the sample as visible in row (I) panels (e-g) of Fig. \ref{frames}. The exact location of these secondary necks is not reproducible in subsequent experiments.
In regime (II) of deformation  (second row from the top in Fig. \ref{frames}) the geometric inhomogeneity of the sample becomes even  more pronounced than in regime (I): above the onset of the primary necking, the diameter of the sample is non constant over the entire length of the sample. Just after a local maximum in the viscosity is observed at $\epsilon_{H} \approx 3.3$, a secondary neck emerges slightly below the center point of the sample, second row, panel (a), Fig. \ref{frames}. 

\begin{figure*}
\begin{center}
\centering
\includegraphics[width=16cm]{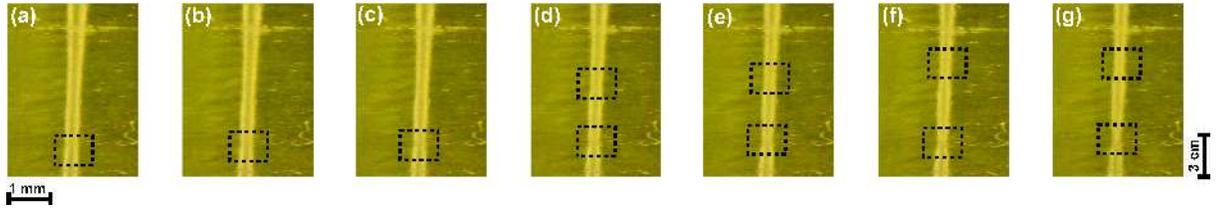}
\caption{Magnified views of the necks highlighted in Fig. \ref{frames} corresponding to regime II (second row).} \label{necks}
\end{center}
\end{figure*}

A magnified view of these necks is presented in Fig. \ref{necks}.

As the Hencky strain increases, the secondary neck becomes sharper (its local diameter decreases rapidly ) and moves slowly along the sample. Another localized neck is formed at $\epsilon_{H} \approx 3.59$ and this ultimately leads to the physical rupture of the sample in a finite time.
The monotonic increase of the error bars during true viscosity measurements within the regimes (I, II) and corresponding to large Hencky strains, Fig. \ref{eta}, can now be easily explained as a result of a systematic increase of sample inhomogeneity due to the emergence of secondary necks.    

 The emergence of secondary necks can also explain the discrepancy between the integral and local transient  extensional viscosity observed within the second regime of deformation (the circles in Fig. \ref{eta} and the inset). Indeed, after the secondary necks are formed along the sample, the integral viscosity measurement which uses a position independent value of the sample diameter, $D_{u}(t)=D_0 exp\left (   -\epsilon_H /2\right )$, systematically overestimates the actual average sample diameter, $D(t)= \langle D(z,t) \rangle_z $. As a consequence, above the onset of the secondary necking, the integral transient elongational viscosity decreases and a viscosity maximum is observed. On the other hand, if the emergence of the secondary necks is accounted for by averaging the stresses along the actual length of the sample, no decrease of viscosity is observed and a rather convincing steady state seems to be reached instead (the inset in Fig. \ref{eta}).

 These experimental findings suggest that the long debated maximum of the transient extensional viscosity does not reflect true rheological features of the material and is solely related to a severe inhomogeneity of deformation states due to the emergence of secondary necks along the sample. This conclusion is consistent with the discussion presented in Sec. \ref{subsec_analytical_condition}: if the viscosity maximum would emerge as a true rheological feature (that is in the absence of geometric inhomogeneities) than, corresponding to this maximum, the tensile force should scale exponentially which, as already discussed above, is not the case within regime (II).

Finally, we turn our attention to the evolution of the sample inhomogeneity during measurements of the transient elongational viscosity in regime (III). Within this deformation regime, the overall homogeneity of the sample is better than within the regimes (I), (II), though curvature effects are visible in the proximity of the plates of the rheometer, row (III), Fig. \ref{frames}. In spite of a better sample homogeneity (no secondary necks are observed in this deformation regime), however, the differences between local and integral measurements of the viscosity are significant (the triangles, Fig. \ref{eta}). This fact deserves a brief explanation. As recently shown in \cite{localelongational2} the relative difference between local and integral measurements of the tensile stress is given by:

\begin{equation}
 \frac{\sigma(t)-\sigma_u(t) }{\sigma_u(t)} =4L^{-1}(t)  \int_0^{L(t)} \frac{\delta(z,t)[D_u(t)+\delta(z,t)]}{[D_u(t)+2\delta(z,t)]^2}dz  \label{eq_error}
\end{equation}
where $\delta(z,t)=\frac{D(z,t)-D_u(t)}{2}$ quantifies the deviation of the sample shape from the ideal cylindrical form and $\sigma_u(t)=\frac{4F(t)}{\pi D^2_u(t)}$ is the integral stress. Assuming $\xi=\delta(z,t)/D_u(t)<1$, it can be easily shown that, to a leading order in $\xi^2$, $\left |  \frac{\sigma(t)-\sigma_u(t) }{\sigma_u(t)} \right | \approx 4\xi \propto exp \left( \dot{\epsilon}/2t\right)$. This explains the increase of the relative stress error with the rate of deformation at a fixed time instant observed in Fig. \ref{eta}.

\subsection{Comparison with results from literature}

In the following, a comparison of our experimental findings with experimental work performed by others is presented. The recent work by Rasmussen et al., \cite{rasmussen} presents a detailed experimental observation of a true viscosity overshoot (i.e. a maximum in the extensional viscosity followed by an extended plateau). Whereas in our experiments a local maximum of the integral elongational viscosity was found during each experiment conducted in region II, a plateau following such maximum has never been observed. 

 In order to clarify the reasons underlying this discrepancy,  we first point out several similarities and differences between our experiments and those reported in \cite{rasmussen}. The material used in both experiments was the same, namely Lupolen 1840 D, and the temperaturea during the experiments were quite comparable: $T=140 ^{\circ} C$ for our experiments and $T=130 ^{\circ} C$ for the experiments reported in \cite{rasmussen}. Within $10$ degrees difference in temperature, we do not expect a significant change in the qualitative behaviour of the transient extensional viscosity. To check this, additional integral viscosity measurements (these findings will be published elsewhere) have been conducted in a wide range of temperatures (from $T=130 ^{\circ} C$ to $T=190 ^{\circ} C$) and a wide range of deformation rates (from $0.001 s^{-1}$ up to $0.3 s^{-1}$) but whenever a viscosity maximum was present it simply led to the physical rupture of the sample without any hint of a plateau regime. Therefore, we rule out the temperature as a decisive factor for the emergence of the viscosity plateau. There are,however, several other differences between the two approaches compared here. Whereas during our experiments the sample under investigation was immersed in an oil bath in order to minimize the buoyancy effects, the experiments presented in \cite{rasmussen} were performed in air and a correction for the gravity effects was employed, \cite{szabo,szabo1}. However, as this correction only subtracts the weight of the sample from the measured tensile force, it cannot be responsible for the "true overshoot" behaviour observed by Rasmussen et al. To sum up the arguments above, we believe that the disagreement between our experimental findings and those reported by Rasmussen et al. has little or nothing to do with the preparation of the samples, the operating temperature and the design of the extensional apparatus. In a last attempt to understand this discrepancy, we compared our data analysis procedure (see the description in Sec. \ref{subsec_anlysis}) with the procedure used by Rasmussen et al, \cite{rasmussen}. There exists a fundamental difference between the two approaches.

\begin{figure*}
\begin{center}
\centering
\includegraphics[width=14cm]{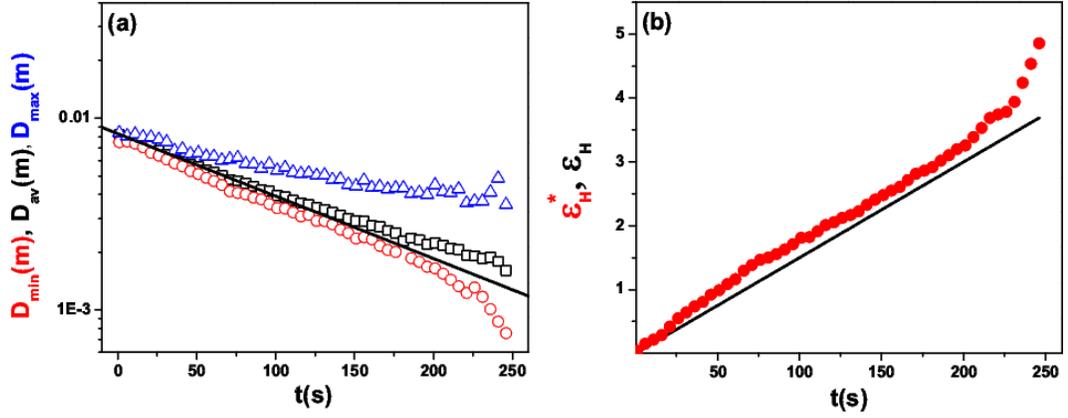}
\caption{(a) Time dependence of the minimum diameter of the sample (circles), the average diameter (squares) and the maximum diameter (triangles). (b) Time dependence of the Hencky strain (calculated using the minimum diameter of the sample, $D_{min}$), $\epsilon^{*}_{~H}$. The full line is the Hencky strain measured using the actual length of the sample, $\epsilon_{~H}$. The data were acquired at constant rate of deformation, $\dot{\epsilon}=0.015~s^{-1}$.} \label{comparison1}
\end{center}
\end{figure*}

Whereas we have defined the Hencky strain using the actual length of the sample, $\epsilon_H(t)=ln \left [ \frac{L(t)}{L_0} \right]$, and measured it accordingly by monitoring the position of the top plate of the rheometer, Rasmussen et al. have defined it as    
$\epsilon^{*}_{H}(t)=-2ln \left [ \frac{D_{mid}(t)}{D_0} \right]$, using the middle plane diameter of the sample, $D_{mid}(t)$. It is obvious that in the case of uniaxial extension at a constant rate of deformation(in time and along the entire sample), the two ways of calculating the Hencky strain are entirely equivalent. In the case of the experiments presented in this paper, however, one clearly deals with a geometrically non-uniform deformation process which ultimately translates into a strong deviation from the idealized uniaxial case. This experimental fact is illustrated in Fig. \ref{frames} where one can clearly see that, within the second deformation regime (the second row from the top), the sample is far from being cylindrical when a viscosity maximum is observed. The impact of the geometric non-uniformity of the sample on the kinematics of the deformation process is illustrated in Fig. \ref{comparison1}.

Above the onset of the primary non-uniformity of the sample (the first maximum of the tensile force which corresponds here to $t \approx 24 s$), the strain becomes strongly localized along the sample. This can be clearly noticed in Fig. \ref{comparison1} (a) where the time dependencies of the minimum sample diameter $D_{min}$, the averaged (along the actual length of the sample) diameter $D_{av}$ and the maximum sample diameter $D_{max}$ are displayed together with the diameter corresponding to a uniform deformation at constant rate (the full line).    
\begin{figure*}
\begin{center}
\centering
\includegraphics[width=8cm]{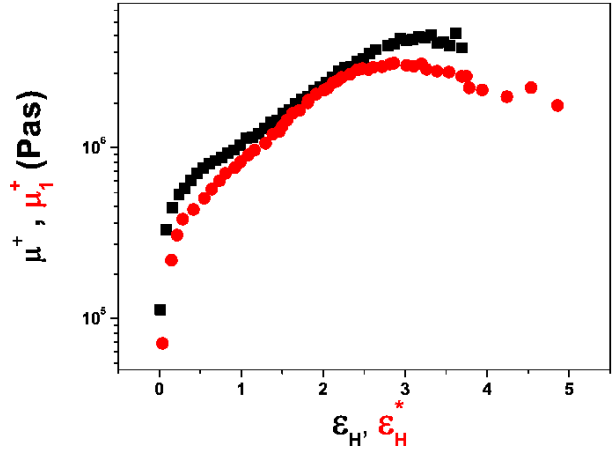}
\caption{Comparison between the transient elongational viscosity measured by our averaging method (squares) and the transient elongational viscosity obtained following the procedure of Rasmussen et al., \cite{rasmussen} (circles).} \label{comparison2}
\end{center}
\end{figure*}
The comparison between the Hencky strains $\epsilon_H$ and $\epsilon^*_{H}$ is presented in Fig. \ref{comparison1} (b).
At late stages of the deformation process (after the viscosity overshoot is observed) the local slope $\frac{dD_{min}(t)}{dt}$ increases drastically suggesting that the highest rate of material deformation corresponds to the necked region of the sample.      
During our experiments, the neck is roughly located around the middle of the sample (though sometimes additional necks may emerge in other places), which is precisely the point where Rasmussen et al. measure the diameter of the sample, \cite{rasmussen}. Although in Ref. \cite{rasmussen} a feedback mechanism has been employed to ensure a constant rate of decay of the mid sample diameter, the assessment of the transient elongational viscosity remains conceptually problematic, because it implies combining an integral quantity (the measured tensile force which reflects the response of the \textit{entire} sample under deformation) with  two locally  measured (around the neck!) kinematic quantities: the strain and  the rate of deformation. 
In order to clearly illustrate this, we analyse in the following our data using the procedure described in Ref. \cite{rasmussen}. The result of such an analysis is presented in Fig. \ref{comparison2} (circles) together with the true extensional viscosity obtained by our method (squares). The extensional viscosity $\mu_{1}^+$ is defined as $\mu_{1}^+(t)=\frac{4F(t)}{\pi \left ( D_{mid}(t) \right )^2 \frac{d \left (\epsilon^{*}_{H}(t) \right)}{dt} }$. One can clearly see that the two data analysis procedures applied to the same raw data (namely the same force signal and the same sequence of sample images) yield strikingly different results. Whereas our procedure hints to a plateau of the transient elongational viscosity, the procedure employed in Ref. \cite{rasmussen} leads to a clear viscosity overshoot behaviour: a viscosity plateau following the viscosity maximum is now visible up the $\epsilon^{*}_H =5$.

As a conclusion, the discrepancy between our transient viscosity measurements and the results presented by Rasmussen et al. originates in the differences between the two approaches: whereas we have used an integral definition for the Hencky strain and averaged the tensile stress along the sample, Ref. \cite{rasmussen} used local values for both the Hencky strain and the stress.   

 \section{Conclusions}
A systematic investigation of the long debated "viscosity overshoot" during the uniaxial extension of a strain hardening polymer melt was presented. 
The mathematical condition for the tensile stress to have a local maximum is presented in Sec. \ref{subsec_analytical_condition}, using no other assumptions except the differentiability of both the tensile force and the tensile stress. According to Eq. \ref{eq_condition}, a local stress maximum   
may be observed during a homogeneous deformation process only if the tensile force scales exponentially around this maximum. If the deformation process is not homogeneous, a stress maximum and an exponential scaling of the tensile force may still be observed if the rates of deformation are small. These theoretical considerations are investigated experimentally corresponding to voarious deformation regimes.   
Depending on the magnitude of the Weissenberg number, we identify three distinct deformation regimes. At low Wi, (regime (I), Fig. \ref{phasediagram}) the integral tensile stress displays a broad maximum, Fig. \ref{forces}(a). In the neighbourhood of the stress maximum, the tensile force decays nearly exponentially with a rate set by the deformation rate, $\dot {\epsilon}$. As within this regime the deformation is inhomogeneous (Fig. \ref{frames}, row I), this nearly exponential scaling can only be explained, according to Eq. \ref{eq_condition}, by the smallness of the deformation rate. The stress maximum observed in regime I should not be confused with the viscosity overshoot phenomenon, which was observed at substantially larger $Wi$, \cite{wagnerovershoot, rasmussen}. We observe such a viscosity maximum for intermediate values of $Wi$, in regime II, Fig. \ref{phasediagram}. This maximum is clearly not consistent with a homogeneous deformation process, because in the neighbourhood of this maximum the tensile force does not scale exponentially, Fig. \ref{forces} (b). As suggested by the convergence of the stress maximum and physical rupture lines (the Hencky strains corresponding to the physical rupture of the sample) visible in Fig. \ref{phasediagram} within regime (II),  the two phenomena are interconnected: the viscosity maximum is just a precursor of the physical rupture of  the sample. Indeed, real time imaging of the sample confirms that right above the stress maximum, secondary necks develop along the sample leading to sample's rupture, Fig. \ref{frames}, row II. Based on the images of the sample, we measure the true tensile stress by averaging the local stresses along the actual length of the sample. Whereas in regime (II) the integral elongational viscosity displays a clear maximum, the true viscosity measurements (which properly account for the presence of necks along the sample) indicate a plateau instead. Therefore we conclude that the viscosity maximum is merely an experimental artefact introduced by the strong geometric inhomogeneity of the sample. In the fast stretching limit (regime III, Fig. \ref{phasediagram}), the homogeneity of the sample is better preserved (Fig. \ref{frames}, row III) and no viscosity maximum is observed, Fig. \ref{forces}(c).    

Finally, our experimental findings are compared with a recent experimental investigation of the viscosity overshoot phenomenon by Rasmussen et al., \cite{rasmussen}. 

The discrepancy between the true extensional viscosity measurements presented in this paper and the results presented in Ref. \cite{rasmussen} is explained by differences in the data analysis procedure. 
As clearly illustrated in Fig. \ref{comparison2}, using the same procedure as in \cite{rasmussen}, one can qualitatively reproduce a viscosity overshoot behaviour as well.

As a final conclusion, neither a local maximum in the transient elongational viscosity nor a true viscosity overshoot behaviour are, according to our study, real rheological features but they only emerge as artefacts due to the strong geometric non-uniformity of the sample at high Hencky strains. The main issue responsible for these artefacts is the geometric inhomogeneity of the sample which becomes critical when secondary necks are formed.   
Existing experimental work on extensional rheology melts in a non-linear range should be reconsidered particularly in relation with the inhomogeneity of sample deformation. Theoretical work should take these findings into account.
\begin{appendix}
Aici trebuie sa adaug calculele
\end{appendix}
%\begin{acknowledgements}
\subsection*{Acknowledgements}
T. B. and Z.S. gratefully acknowledge the financial support from the German
Research Foundation (grants $MU 1336/6-4$ and $STA 1096/1-1$,
respectively). T. B. and Z.S. thank Mrs. Magdalena Papp for her assistance during some of the experiments presented in this study. One of us (T.B.) thanks Alfred Frey for valuable technical advice, assistance with the M\"{u}nstedt rheometer, and for the implementation of the digital trigger for the camera.
%\end{acknowledgements}

%\bibliographystyle{elsarticle-num-names.bst}

\bibliographystyle{apalike}
%\bibliography{biblio}

\begin{thebibliography}{}

\bibitem[Boukany and Wang, 2009]{wang1}
Boukany, P.~E. and Wang, S.~Q. (2009).
\newblock Universal scaling behavior in startup shear of entangled linear
  polymer melts.
\newblock {\em Journal of Rheology}, 53:617.

\bibitem[Burghelea and M\"{u}nstedt, 2009]{localelongational2}
Burghelea, T. and M\"{u}nstedt, H. (2009).
\newblock Influence of sample inhomogeneity on the linear elongational
  viscosities of two low density polyethylenes.
\newblock {\em submitted to Applied Rheology}.

\bibitem[Burghelea et~al., 2009]{localelongational1}
Burghelea, T., Star\'{y}, Z., and M\"{u}nstedt, H. (2009).
\newblock Local versus global measurements of the extensional viscosity.
\newblock {\em accepted for publication in Journal of Rheology}.

\bibitem[Consid\`{e}re, 1885]{considere}
Consid\`{e}re, A. (1885).
\newblock {\em Annales des Ponts et Chauss\'{e}es}, 6(9):574.

\bibitem[Hepperle and M\"{u}nstedt, 2005]{m2}
Hepperle, J. and M\"{u}nstedt, H. (2005).
\newblock Rheological properties of branched polystyrenes: nonlinear shear and
  extensional behavior.
\newblock {\em Rheologica Acta}, 45:717--727.

\bibitem[Joshi and Denn, 2003]{denn2003}
Joshi, D.~M. and Denn, M. (2003).
\newblock Rupture of entangled polymer liquids in elongational flow.
\newblock {\em Journal of Rheology}, 47:291--298.

\bibitem[Joshi and Denn, 2004]{denn2004}
Joshi, D.~M. and Denn, M. (2004).
\newblock Failure and recovery of entangled polymer melts.
\newblock {\em In: Binding D. M., Walters K., eds. Rheology Reviews,
  Aberystwyth: British Society of Rheology}.

\bibitem[Lyhne et~al., 2009]{hassager2009PRL}
Lyhne, A., Rasmussen, Henrik, K., and Hassager, O. (2009).
\newblock Simulation of elastic rupture in extension of entangled monodisperse
  polymer melts.
\newblock {\em Physical Review Letters}, 102:138301.

\bibitem[McLeish and Larson, 1998]{mcleish}
McLeish, T. C.~B. and Larson, R.~G. (1998).
\newblock Molecular constitutive equations for a class of branched polymers:
  The pom-pom polymer.
\newblock {\em Journal of Rheology}, 42:81--110.

\bibitem[Meissner and Hostettler, 1994]{meissner1}
Meissner, J. and Hostettler, J. (1994).
\newblock A new elongational rheometer for polymer melts and other highly
  viscous liquids.
\newblock {\em Rheologica Acta}, 33:1--21.

\bibitem[M\"{u}nstedt, 1979]{m4}
M\"{u}nstedt, H. (1979).
\newblock New universal extensional rheometer for polymer melts. \
  {M}easurements on a polystyrene sample.
\newblock {\em Journal of Rheology}, 23:421--436.

\bibitem[M\"{u}nstedt et~al., 1998]{m1}
M\"{u}nstedt, H., Kurzbeck, S., and Egersd\"{o}rfer, L. (1998).
\newblock Influence of molecular structure on rheological properties of
  polyethylenes.
\newblock {\em Rheologica Acta}, 37(1):21--29.

\bibitem[Nielsen et~al., 2006]{nielsen1}
Nielsen, J.~K., Rasmussen, H.~K., Hassager, O., and McKinley, G.~H. (2006).
\newblock Elongational viscosity of monodisperse and bidisperse polystyrene
  melts.
\newblock {\em Journal of Rheology}, 50:453--476.

\bibitem[Nordmaier et~al., 1990a]{nordmaier1}
Nordmaier, E., Lanver, U., and Lechner, M.~D. (1990a).
\newblock The molecular structure of low-density polyethylene 1. long-chain
  branching and solution properties.
\newblock {\em Macromolecules}, 23:1072--1076.

\bibitem[Nordmaier et~al., 1990b]{nordmaier2}
Nordmaier, E., Lanver, U., and Lechner, M.~D. (1990b).
\newblock The molecular structure of low-density polyethylene 2. particle
  scattering factors.
\newblock {\em Macromolecules}, 23:1077--1084.

\bibitem[Raible et~al., 1979]{meissnerovershoot}
Raible, T., Demarmels, A., and Meissner, J. (1979).
\newblock Stress and recovery maxima in ldpe melt elongation.
\newblock {\em Polymer Bulletin}, 1:397--402.

\bibitem[Rasmussen et~al., 2005]{rasmussen}
Rasmussen, H.~K., Nielsen, J.~K., Bach, A., and Hassager, O. (2005).
\newblock Viscosity overshoot in the start-up of uniaxial elongation of low
  density polyethylene melts.
\newblock {\em Journal of Rheology}, 49(2):369--381.

\bibitem[Ravindranath and Wang, 2008]{wang4}
Ravindranath, S. and Wang, S.-Q. (2008).
\newblock Universal scaling characteristics of stress overshoot in startup
  shear of entangled polymer solutions.
\newblock {\em Journal of Rheology}, 99:681--695.

\bibitem[Resch et~al., 2009]{resch1}
Resch, J.~A., Stadler, F.~J., Kaschta, J., and M\"{u}nstedt, H. (2009).
\newblock Temperature dependence of the linear steady-state shear compliance of
  linear and long-chain branched polyethylens.
\newblock {\em Macromolecules}, 42:5676--5683.

\bibitem[Schweizer, 2000]{schweizer}
Schweizer, T. (2000).
\newblock The uniaxial elongational rheometer rme - six years of experience.
\newblock {\em Rheologica Acta}, 39(5):428--443.

\bibitem[Sentmanat, 2003a]{sentamat2}
Sentmanat, M.~L. (2003a).
\newblock Dual wind up extensional rheometer.
\newblock {\em US Patent No. 6578413}.

\bibitem[Sentmanat, 2003b]{sentamat1}
Sentmanat, M.~L. (2003b).
\newblock A novel device for characterizing polymer flows in uniaxial
  extension.
\newblock {\em Soc. Plastics Engineers, Tech. Papers, CD-ROM}, 49.

\bibitem[Strobl, 2007]{strobl}
Strobl, G. (2007).
\newblock {\em The physics of polymers}.
\newblock Springer-Verlag Berlin Heidelberg.

\bibitem[Szabo, 1997]{szabo}
Szabo, P. (1997).
\newblock Transient filament stretching rheometer part i: Force balance
  analysis.
\newblock {\em Rheologica Acta}, 36:277--284.

\bibitem[Szabo and McKinley, 2003]{szabo1}
Szabo, P. and McKinley, Gareth, H. (2003).
\newblock Filament stretching rheometer: Inertia compensation revisited.
\newblock {\em Rheologica Acta}, 42:269--272.

\bibitem[Tropea et~al., 2007]{handbook}
Tropea, C., Yarin, A.~L., and Foss, J.~S. (2007).
\newblock {\em Handbook of experimental fluid dynamics}.
\newblock Springer-Verlag, Berlin Heidelberg.

\bibitem[Wagner et~al., 1979]{wagnerovershoot}
Wagner, M.~H., Raible, T., and Meissner, J. (1979).
\newblock Tensile stress overshoot in uniaxial extension of ldpe melt.
\newblock {\em Rheologica Acta}, 18:427--428.

\bibitem[Wagner and Rol\'{o}n-Garrido, 2008]{wagner3}
Wagner, M.~H. and Rol\'{o}n-Garrido, V.~H. (2008).
\newblock Verification of branch point withdrawal in elongational flow of
  pom-pom polystyrene melt.
\newblock {\em Journal of Rheology}, 52(5):1049--1068.

\bibitem[Wagner et~al., 2001]{wagner4}
Wagner, M.~H., Rubio, P., and Bastian, H. (2001).
\newblock The molecular stress function model for polydisperse polymer melts
  with disspative convective constraint release.
\newblock {\em Journal of Rheology}, 45:1387--1412.

\bibitem[Wang, 2008]{wang2}
Wang, S.~Q. (2008).
\newblock The tip of iceberg in nonlinear polymer rheology: Entangled liquids
  are "solids".
\newblock {\em Journal of Polymer Science Part B: Polymer Physics},
  46:2660--2665.

\bibitem[Wang et~al., 2007]{wang3}
Wang, Y., Boukany, P., Wang, S.-Q., and Wang, X. (2007).
\newblock Elastic breakup in uniaxial extension of entangled polymer melts.
\newblock {\em Physical Review Letters}, 99:237801.

\bibitem[Yu. et~al., 2009]{hassager2009}
Yu., K., Mar\'{i}n, J. M.~R., and Hassager, O. (2009).
\newblock 3d modeling of dual wind-up extensional rheometers.
\newblock {\em accepted for publication in Journal of Non-Newtonian Fluid
  Mechanics}.

\end{thebibliography}

\newpage
\section*{List of Figures}

\begin{itemize}

%f1
\item{\textbf{Fig.1}: (a) Schematic view of the experimental apparatus: \textbf{C}- oil bath, $\mathbf{P_{1}}$ and $\mathbf{P_{2}}$ - top and bottom plates of the rheometer, \textbf{S}- the sample under investigation, \textbf{M}- AC servo motor, \textbf{D}- the control drive of the rheometer, $\mathbf{PC_{1,2}}$ - personal computers, \textbf{TL}- telecentric lens, \textbf{CCD}- video camera. (b) Sample illumination and imaging: $\mathbf{LS_{1}}$ and $\mathbf{LS_{2}}$- linear light sources, \textbf{S}- the sample under investigation. (c)  Example of a telecentric sample image corresponding to $\epsilon_{H}=2.7$. The field of view was actually larger but the image has been cropped for clarity reasons.  (d) Principle of the local measurements of the extensional viscosity. The vertical dotted lines represent the contour of an ideal uniform sample.}

%f2
\item{\textbf{Fig.2}: Transient elongational viscosities at various rates of deformation: (1)- $\dot{\epsilon= 0.002 s^{-1}}$, (2)- $\dot{\epsilon=0.015 s^{-1}}$, (3)- $\dot{\epsilon=0.02 s^{-1}}$, (4)- $\dot{\epsilon= 0.025 s^{-1}}$, (5)- $\dot{\epsilon=0.09 s^{-1}}$. Each data set has been acquired until the physical rupture of the sample occurred.}

%f3
\item{\textbf{Fig.3}: Transient tensile forces and stresses corresponding to different regimes of extension: (a) $\dot{\epsilon} =0.001 s^{-1} (Wi=1.1)$ ,
(b) $\dot{\epsilon} =0.05 s^{-1} (Wi=55)$, (b) $\dot{\epsilon} =0.3 s^{-1} (Wi=330)$.
The inset in panel (a) displays the image of the sample corresponding to the stress maximum.
}

%f4
\item{\textbf{Fig.4}: Dependence of the Hencky strain corresponding to the stress maximum (squares) and physical rupture of the sample (circles) on the Weisenberg number, $Wi$. The vertical dotted lines delineate the extension regimes (I), (II) and (III).}

%f5
\item{\textbf{Fig.5}: Comparison between the integral (full lines) and the locally measured elongational viscosity in each of the deformation regimes presented in Fig. \ref{phasediagram}: squares - $Wi= 2.2$ (regime I), circles - $Wi=16.5$ (regime II), triangles - $Wi=300$ (regime III). The error bars are defined by the root mean square deviation of the local stresses along the sample. The inset presents a magnified view of the data acquired in regime II. The vertical arrows indicate the physical rupture of the sample.}

%f6
\item{\textbf{Fig.6}: Sequence of specimen images under deformation at different Wi. The image rows (from top to bottom) correspond to: $Wi= 2.2$ (region I), $Wi=16.5$ (region II), $Wi=99$ (region III). The aspect ratio of each image has been modified in order to enhance the clarity. The dotted squares indicate the location of the necks. The Hencky strains are indicated on the top of each image.}

%f7
\item{\textbf{Fig.7}: Magnified views of the necks highlighted in Fig. \ref{frames} corresponding to regime II (second row).}

%f8
\item{\textbf{Fig.8}: Time dependence of the minimum diameter of the sample (circles), the average diameter (squares) and the maximum diameter (triangles). (b) Time dependence of the Hencky strain (calculated using the minimum diameter of the sample, $D_{min}$), $\epsilon^{*}_{~H}$. The full line is the Hencky strain measured using the actual length of the sample, $\epsilon_{~H}$. The data were acquired at constant rate of deformation, $\dot{\epsilon}=0.015~s^{-1}$.}

%f9
\item{\textbf{Fig.9}: Comparison between the transient elongational viscosity measured by our averaging method (squares) and the transient elongational viscosity obtained following the procedure of Rasmussen et al., \cite{rasmussen} (circles).}

\end{itemize}

\end{document}